\documentclass[onecolumn,preprint]{revtex4-1}

\usepackage{amsmath,amsfonts,amssymb}
\usepackage{graphicx}
\usepackage{dcolumn}
\usepackage{bm}
\usepackage{xcolor}

\definecolor{arsenic}{rgb}{0.23, 0.27, 0.29}
\definecolor{blue-violet}{rgb}{0.54, 0.17, 0.89}
\definecolor{britishracinggreen}{rgb}{0.0, 0.26, 0.15}
\definecolor{ferngreen}{rgb}{0.31, 0.47, 0.26}
\definecolor{forestgreen}{rgb}{0.13, 0.55, 0.13}
\definecolor{limegreen}{rgb}{0.2, 0.8, 0.2}

\newcommand{\ket}[1]{\vert\mskip3mu#1\mskip3mu\rangle}
\newcommand{\braket}[2]{\langle\mskip3mu#1\mskip3mu\ket{#2}}
\newcommand{\braopket}[3]{\langle\mskip3mu#1\mskip3mu\vert\mskip3mu#2\mskip3mu\ket{#3}}

\newcommand{\bmS}{{\bm S}}
\newcommand{\bmK}{{\bm K}}
\newcommand{\bmQ}{{\bm Q}}
\newcommand{\bmF}{{\bm F}}
\newcommand{\bmG}{{\bm G}}

\newcommand{\bmI}{{\bm I}}
\newcommand{\bmT}{{\bm T}}
\newcommand{\bmU}{{\bm U}}

\newcommand{\bmW}{{\bm W}}
\newcommand{\bmB}{{\bm B}}
\newcommand{\oOmega}{\overline{\Omega}}

\begin{document}

\title{Efficient computational methods for rovibrational
transition rates in molecular collisions}

\author{Taha Selim} 
\email{tselim@science.ru.nl}
\affiliation{Theoretical Chemistry \\
Institute for Molecules and Materials, Radboud University \\
Heyendaalseweg 135, 6525 AJ Nijmegen, The Netherlands}

\author{Ad van der Avoird} 
\affiliation{Theoretical Chemistry \\
Institute for Molecules and Materials, Radboud University \\
Heyendaalseweg 135, 6525 AJ Nijmegen, The Netherlands}

\author{Gerrit C. Groenenboom}  
\email{gerritg@theochem.ru.nl}
\affiliation{Theoretical Chemistry \\
Institute for Molecules and Materials, Radboud University \\
Heyendaalseweg 135, 6525 AJ Nijmegen, The Netherlands}

\date{\today}

\begin{abstract}
Astrophysical modeling of processes in environments that are not in
local thermal equilibrium requires the knowledge of state-to-state rate
coefficients of rovibrational transitions in molecular collisions. These
rate coefficients can be obtained from coupled-channel (CC) quantum
scattering calculations which are very demanding, however. Here we
present various approximate, but more efficient methods based on the
coupled-states approximation (CSA) which neglects the off-diagonal
Coriolis coupling in the scattering Hamiltonian in body-fixed
coordinates. In particular, we investigated a method called NNCC
(nearest-neighbor Coriolis coupling) [D.~Yang, X.~Hu, D.~H.~Zhang,
and D.~Xie, J. Chem. Phys. {\bf 148}, 084101 (2018)] that includes
Coriolis coupling to first order. The NNCC method is more
demanding than the common CSA method, but still much more efficient
than full CC calculations, and it is substantially more accurate than
CSA. All of this is illustrated by showing state-to-state cross sections
and rate coefficients of rovibrational transitions induced in CO$_2$ by
collisions with He atoms. It is also shown that a further reduction of
CPU time, practically without loss of accuracy, can be obtained by
combining the NNCC method with the multi-channel distorted-wave Born
approximation (MC-DWBA) that we applied in full CC calculations in a
previous paper.
\end{abstract}

\maketitle

\section{Introduction}
\label{sec:intro}
In modeling protoplanetary disks and other interstellar media that are
not in local thermal equilibrium (LTE) with the aid of spectroscopic
data from ground and satellite based telescopes
\cite{pontoppidan:08,mandell:12,bruderer:15,bosman:17,bosman:19}, the
effects of molecular collisions are important. In such non-LTE
environments the populations of the rovibrational states of a molecule
---and thereby the characteristics of its spectrum--- are not only
determined by absorption and emission of electromagnetic radiation, but
also by transitions induced by molecular collisions. By analyzing these
spectra one gets crucial information not only about the abundance of
various molecules, but also about the local conditions. An essential
element in this analysis is the knowledge of the rate coefficients of
rovibrational transitions induced by the collisions of the molecule with
H$_2$ molecules, He atoms, and electrons. Quantum scattering
calculations, based on intermolecular potentials determined by
\textit{ab initio} electronic-structure calculations, can provide
inelastic collision cross sections, from which the required transition
rate coefficients and their temperature dependence can be derived.

An important molecule in these studies is carbon dioxide, CO$_2$
\cite{oberg:11,bosman:17,bosman:19}. Accurate cross sections and rate
coefficients for rotationally inelastic CO$_2$-He collisions were
recently reported by Godard Palluet \textit{et al.} \cite{palluet:22}.
CO$_2$ has no permanent dipole moment, so its rotational transitions are
forbidden and it cannot be observed in microwave or far-infrared
spectra. But it can be observed in mid- and near-infrared spectra
through its vibrational transitions. It has three vibrational modes: a
twofold degenerate bend mode with experimental frequency 667~cm$^{-1}$,
an asymmetric stretch mode at 2349~cm$^{-1}$ , and a symmetric stretch
mode at 1333~cm$^{-1}$. The latter mode is not infrared active by itself
but becomes observable through a Fermi resonance with the bend overtone.
In the pioneering theoretical studies of rate coefficients for
vibrational transitions in CO$_2$ induced by collisions with rare gas
(Rg) atoms by Clary \textit{et al.}
\cite{clary:81,clary:82,banks:87b,wickham:87b} they used VCC-IOS, a
vibrational coupled-channel (CC) method for the vibrations, combined
with the infinite-order sudden (IOS) approximation for the rotations.
This method provided rate coefficients for vibrational transitions,
without considering specific initial and final rotational states. The
more advanced models currently being developed by astronomers
\cite{bosman:17,bosman:19} and the availability of data from the James
Webb space telescope (JWST) in the near future require rovibrational
state-to-state collisional rate coefficients. These can nowadays be
obtained from the numerically exact coupled-channel (CC) method with the
use of accurate \textit{ab initio} calculated intermolecular potentials.
Full CC calculations are still time-consuming, however, especially at
the higher collision energies needed to obtain rate coefficients for
higher temperatures. In a previous paper \cite{selim:21} we have shown
how one can reach the CC level of accuracy with a less time-consuming
procedure that handles the coupling between rotational states by the CC
method and the weaker coupling between different vibrational states with
the multichannel distorted-wave Born approximation (MC-DWBA). Here we
investigate further possibilities to speed up the calculation of
collisional rate coefficients for rovibrational transitions by using the
coupled-states approximation (CSA) and an improvement of it that
includes Coriolis coupling to first order. We apply various methods to
CO$_2$-He collisions with CO$_2$ excited in the symmetric stretch mode.
We consider both the efficiency of these methods and the accuracy of the
results they provide.

\section{Theory}
\label{sec:theory}
\subsection{Coupled channels method}
\label{sec:CC}
The methods discussed in the present paper are based on the
coupled-channels (CC) ---also called close-coupling--- method,
formulated in body-fixed (BF) coordinates. These coordinates refer to a
BF frame with its $z$-axis along the vector $\bm{R}$ that points from
the center of mass of CO$_2$ to the He nucleus and the CO$_2$-He complex
lying in the $xz$-plane. The coordinates are the length $R$ of the
vector $\bm{R}$, the angle $\theta$ between the CO$_2$ axis and the
vector $\bm{R}$, and the normal coordinate $Q$ along which CO$_2$ is
deformed with respect to its linear equilibrium geometry with
equal C-O bond lengths of 1.162~\AA.

The 3D Hamiltonian of vibrotor-atom system CO$_2$-He over a
monomer normal coordinate $Q$ is given, in the BF frame, by
\begin{equation}
\label{eq:hamiltonian}
\hat{H} =  - \frac{\hbar^2}{2 \mu R} \frac{\partial^{2}}{\partial R^{2}} R
 + \hat{H}_{\textrm{CO}_2}(Q)
 + \frac{\hat{J}^2 + \hat{j}^2 - 2\hat{\bm{j}} \cdot \hat{\bm{J}}}{2 \mu R^{2}}
  + V(Q,R,\theta),
\end{equation}
where $\mu =
m_{\textrm{CO}_2}m_{\textrm{He}}/(m_{\textrm{CO}_2}+m_{\textrm{He}})$ is
the reduced mass of the complex, $\hat{\bm{j}}$ the CO$_2$ monomer
rotational angular momentum operator, $\hat{\bm{J}}$ the total angular
momentum operator of the complex, and $\hat{J}^2 + \hat{j}^2 -
2\hat{\bm{j}} \cdot \hat{\bm{J}}$ represents the end-over-end angular
momentum operator $L^{2}$ in the BF-frame \cite{avoird:94}. The monomer
Hamiltonian $\hat{H}_{\textrm{CO}_2}(Q)$ is defined in Eq.~(2) of
Ref.~\cite{selim:21} and also the computation of the normal modes $Q$
---its eigenstates in the rigid-rotor harmonic-oscillator
approximation--- is described there. Here we consider the
symmetric stretch mode $Q \equiv Q_1 = 0.17678 \Delta z_1 - 0.17678
\Delta z_3$ $a_0$, where $\Delta z_1$ and $\Delta z_3$ are the
displacements of the O atoms along the CO$_2$ axis with respect to
their equilibrium positions. The symmetric stretch mode does not
displace the C atom.

The eigenfunctions of the CO$_2$ monomer Hamiltonian are
\begin{equation}
\label{eq:monbasis}
 \ket{v j \Omega} = \chi_{vj}(Q) Y_{j\Omega}(\theta,\phi)
\end{equation}
and the corresponding eigenvalues are $\epsilon_{vj}$.
The vibrational functions $\chi_{vj}(Q)$ are $j$-dependent,
the rotational functions $Y_{j\Omega}(\theta,\phi)$ are spherical harmonics.
The latter are expressed with respect to the BF frame, the angle
$\theta$ is the same as defined above and the angle $\phi$ coincides
with the third Euler angle for the overall rotation of the complex.
The quantum number $\Omega$ is the projection of the CO$_2$ angular
momentum $\hat{\bm{j}}$ and of the total angular momentum
$\hat{\bm{J}}$ on the BF $z$-axis along the vector $\bm{R}$.

The channel basis in coupled channels (CC) scattering calculations for
rovibrationally inelastic CO$_2$-He collisions is, in BF coordinates,
\begin{equation}
\label{eq:basis}
\ket{v j \Omega; J M_J} = \sqrt{\frac{2J+1}{4\pi}} \chi_{vj}(Q)
   Y_{j\Omega}(\theta,0) D^{J}_{M_J \Omega}(\alpha,\beta,\phi)^* .
\end{equation}
The angles $(\beta,\alpha)$ are the polar angles of the vector $\bm{R}$
with respect to a space-fixed frame (SF), and the Euler angles
$(\alpha,\beta,\phi)$ in the Wigner $D$-functions describe the orientation
of the BF frame relative to the SF frame. The angle $\phi$ has been
moved from the spherical harmonics in Eq.~(\ref{eq:monbasis}) to the
overall rotation functions, which is mathematically equivalent.

The Coriolis coupling operator $2\hat{\bm{j}} \cdot \hat{\bm{J}}$ in
Eq.~(\ref{eq:hamiltonian}) can be written as
\begin{equation}
\label{eq:coriolis}
2\hat{\bm{j}} \cdot \hat{\bm{J}} = 2\hat{j}_z \hat{J}_z + \hat{j}_+
\hat{J}_+ + \hat{j}_- \hat{J}_- .
\end{equation}
The so-called helicity quantum number $\Omega$ is an eigenvalue of both
$\hat{j}_z$ and $\hat{J}_z$. It is an approximate quantum
number; basis functions with different $\Omega$ are mixed by the ladder
operators $\hat{j}_\pm \hat{J}_\pm$ in Eq.~(\ref{eq:coriolis}) which
couple functions with $\Omega$ to those with $\Omega\pm 1$.

Also the \textit{ab initio} calculation of the 3D CO$_2$-He potential
with CO$_2$ deformed along the symmetric stretch coordinate $Q_1$ is
described in Ref.~\cite{selim:21}. The well depth of this potential for
CO$_2$ at its equilibrium geometry is 47.43~cm$^{-1}$. When this potential
is expanded in Legendre polynomials $P_{\lambda}(\cos \theta)$ of order
$\lambda$, as in Ref.~\cite{selim:21}
\begin{equation}
\label{eq:Vpot_3D_eval}
V(Q,R,\theta) = \sum_{\lambda} C_{\lambda}(Q,R) P_{\lambda}(\cos \theta),
\end{equation}
its matrix elements over the BF basis are
\begin{eqnarray}
\label{eq:Vmat}
V_{v' j' \Omega';v j \Omega}(R) & = &
\braopket{v' j'\Omega'; J M_J}{V(Q,R,\theta)}{v j\Omega; J M_J}
\nonumber \\
& = & \delta_{\Omega' \Omega} \; \sum_{\lambda} (-1)^{\Omega'}
[(2j'+1)(2j+1)]^{1/2}
\begin{pmatrix}
j' & \lambda & j \\
0 &    0    &  0
\end{pmatrix}
\begin{pmatrix}
       j' & \lambda & j \\
  -\Omega &    0    &  \Omega
\end{pmatrix}
\nonumber \\
& & \times \braopket{v'j'(Q)}{C_{\lambda}(Q,R)}{v j(Q)}.
\end{eqnarray}
Equation~(\ref{eq:Vmat}) shows the advantages of the BF basis:
the potential $V(Q,R,\theta)$ does not couple functions with different
$\Omega$ and the expression for its remaining matrix elements is simpler
than in the SF basis \cite{arthurs:60}, which makes the calculations
more efficient.

The overall angular momentum $J$ and its projection $M_J$ on the SF
$z$-axis are exact quantum numbers and also the overall parity $P$ under
inversion of the system is a conserved quantity. The basis in
Eq.~(\ref{eq:basis}) is not invariant under inversion; a parity adapted
basis is
\begin{equation}
\label{eq:symbas}
\ket{v j \tilde{\Omega}; P J M_J} =
\left[\ket{v j\,\tilde{\Omega}; P J M_J} + P (-1)^{J}
\ket{v j\, {-\tilde{\Omega}}; P J M_J}\right]/\sqrt{2(1+\delta_{\tilde{\Omega}\,0})},
\end{equation}
where $\tilde{\Omega} \ge 0$ and $P = \pm 1$ is the overall parity.

Another valid symmetry operation is the interchange $P_{13}$ of the O
atoms in CO$_2$. This operator affects only the monomer wave functions
in the basis of Eq.~(\ref{eq:basis}). For the symmetric stretch mode
that we consider here, we find
\begin{equation}
\label{eq:p12}
\hat{P}_{13} \ket{v j \tilde{\Omega}} = (-1)^{j} \ket{v j \tilde{\Omega}}
\end{equation}
Since $^{16}$O nuclei are bosons with spin zero, the wave functions must
be symmetric under $\hat{P}_{13}$. This implies that only functions with
even $j$ are allowed.

The scattering wave functions in CC calculations are written in terms of
the parity-adapted BF channel basis as
\begin{equation}
  \Psi^{P J M_J} = \frac{1}{R}\sum_{v j \tilde{\Omega}}
  \ket{v j \tilde{\Omega}; P J M_J} \psi^{P J M_J}_{v j \tilde{\Omega}}(R).
\end{equation}
When these functions are substituted into the time-independent
Schr{\"o}dinger equation, it follows that the radial wave functions
$\psi^{P J}_{v j \tilde{\Omega}}(R)$ must obey a set of coupled second
order differential equations, the CC equations
\begin{equation}
\label{eq:CC}
  \frac{\partial^{2} }{\partial R^{2}} \psi^{P J}_{v'j'\tilde{\Omega}'}(R) =
   \sum_{v j\tilde{\Omega}} W_{v' j' \Omega'; v j \Omega}^{P J}(R)
    \psi_{vj\tilde{\Omega}}^{P J}(R),
\end{equation}
or in matrix form
\begin{equation}
\label{eq:CCmat}
 \bm{\psi}''(R) = \bmW(R) \bm{\psi}(R).
\end{equation}
The column vector $\bm{\psi}(R)$ contains the radial wave
functions $\psi^{P J}_{v j\tilde{\Omega}}(R)$.
The quantum number $M_J$ has been omitted, since the solutions do not
depend on it. The elements of the matrix $\bmW$ over the primitive
basis in Eq.~(\ref{eq:basis}) are given by
\begin{equation}
\label{eq:Wmat}
W_{v' j' \Omega'; v j \Omega}^{J}(R) =
-\delta_{v'v}\delta_{j'j}\delta_{\Omega' \Omega} k_{vj}^2
      +  T_{v' j' \Omega'; v j \Omega}^{J}(R)
      + 2 \mu V_{v' j' \Omega'; v j \Omega}(R) ,
\end{equation}
with
\begin{equation}
k_{vj}^2 = 2\mu (E - \epsilon_{vj}),
\end{equation}
and $E$ being the total energy. The matrix elements of the potential are
defined in Eq.~(\ref{eq:Vmat}); this matrix is diagonal in $\Omega$.
Only the matrix $\bmT$ which originates from the angular kinetic
energy operator is not diagonal in $\Omega$. Its elements are
\begin{eqnarray}
\label{eq:T}
  T_{v' j' \Omega'; v j \Omega}^{J}(R) =
    && R^{-2} \delta_{v'v}\delta_{j'j}
    \Big\{ \delta_{\Omega' \Omega} \left[J(J+1 + j(j+1) - 2\Omega^2\right] \Big.
\nonumber \\
  && \Big. -\delta_{\Omega' \Omega\pm 1}
   \left[J(J+1-\Omega(\Omega \pm 1)\right]^{1/2}
   \left[j(j+1-\Omega(\Omega \pm 1)\right]^{1/2} \Big\}
\end{eqnarray}
So $\bmT$, and therefore also $\bmW$, contains a series of blocks
diagonal in $\Omega$ and a series of neighboring blocks with $\Omega' =
\Omega \pm 1$. All other elements of these matrices are zero, thanks to
the use of a BF basis in which the potential matrix $\bm{V}$ is
diagonal in $\Omega$.

We solve these equations with the renormalized Numerov propagator method
\cite{johnson:78,johnson:79}. This method implies that one defines an
equidistant grid $R_i, i = 1,\ldots,n$ and propagates the matrix $\bmQ$,
which defines the ratio of the radial wave functions in subsequent
grid points $R_{i-1}$ and $R_i$
\begin{equation}
\bm{\psi}(R_{i-1}) = \bmQ_i \bm{\psi}(R_i) .
\end{equation}
The propagation starts at small $R_1$, where the potential is
sufficiently repulsive that the wave function ---and therefore
$\bmQ$--- is zero, and continues to large $R_n$, where the potential
has vanished. Then, we assume that the radial wave functions
obey flux-normalized $K$-matrix boundary conditions at large $R$
\begin{equation}
\label{eq:psi_asymptotic}
   \bm{\psi}(R) = \bmF(R) -  \bmG(R) \bmK.
\end{equation}
The symbol $\bm{\psi}(R)$ is here used for a matrix with column
vectors that are the solutions of the CC equations in
Eq.~(\ref{eq:CCmat}). The blocks of the matrices $\bmF(R)$ and
$\bmG(R)$ for the open channels are diagonal
\begin{eqnarray}
\label{eq:match}
  F_{v'j'L';vjL}(R) &=& \delta_{v'j';vj} \; \delta_{L'L} \;
                       k_{vj}^{1/2}\, R \, j_{L}(k_{vj} R)
\nonumber \\
  G_{v'j'L';vjL}(R) &=& \delta_{v'j';vj} \; \delta_{L'L} \;
                       k_{vj}^{1/2}\, R \, y_{L}(k_{vj} R),
\end{eqnarray}
and contain asymptotic wave functions that are proportional to
spherical Riccati-Bessel functions \cite{abramowitz:64} of the
first and second kind $j_{L}(z)=\sqrt{\frac{1}{2} \pi /z}
J_{L+\frac{1}{2}}(z)$ and $y_{L}(z)=\sqrt{\frac{1}{2} \pi /z}
Y_{L+\frac{1}{2}}(z)$. Similarly, closed channels are matched to
modified spherical Bessel functions of the first and second kind
$I_{L+\frac{1}{2}}(z)$ and $K_{L+\frac{1}{2}}(z)$, respectively, which
occur in off-diagonal blocks of the matrices $\bmF(R)$ and $\bmG(R)$.

The asymptotic wave functions are defined in the SF frame and
depend on the partial wave index $L$, while the matrix $\bmQ_n$ is
obtained from the propagation in BF coordinates. Therefore, this matrix
is first transformed to SF coordinates
\begin{equation}
\label{eq:transQ}
\bmQ^\textrm{SF}_n = \bmU^\dagger \bmQ_n \bmU.
\end{equation}
The elements of $\bmU$
\begin{equation}
\label{eq:U}
  U^{J j}_{\Omega L} =  \braket{j\Omega L 0}{J\Omega}\sqrt{\frac{2L+1}{2J+1}}
\end{equation}
contain Clebsch-Gordan coefficients $\braket{. . . .}{..}$ \cite{selim:21}.
The matrix $\bmK$ can then be obtained from $\bmQ^\textrm{SF}_n$ by
solving the linear equations
\begin{equation}
\label{eq:matK}
\Big[\bmG(R_{n-1})-\bmQ^\textrm{SF}_n \bmG(R_n)\Big]\bmK
    = \bmF(R_{n-1})-\bmQ^\textrm{SF}_n \bmF(R_n).
\end{equation}
Finally, we use the open-channel block $\bm{K}_{\rm oo}$ of
the matrix $\bm{K}$ to obtain the scattering matrix
\begin{equation}
 \label{eq:S-matrix}
\bmS = (\bmI - i\bmK_{\rm oo})^{-1} (I + i\bmK_{\rm oo}),
\end{equation}
with $\bmI$ being the unit matrix,
and compute state-to-state scattering cross sections
\begin{equation}
\label{eq:ICS}
\sigma_{v',j' \leftarrow v,j}(E) = \frac{\pi}{(2{j}+1) k^{2}_{vj}}
   \sum_{P\,J} (2J+1) \sum_{L'=|J-j'|}^{J+j'}
    \sum_{L=|J-j|}^{J+j} \left|
    \delta_{v'v} \delta_{j'j}\delta_{L'L} -
  S^{P\,J}_{v'j'L';vjL}(E)\right|^{2}.
\end{equation}
Expressions for the temperature ($T$) dependent rate coefficients
$k_{v',j' \leftarrow v,j}(T)$ of the transitions from rovibrational
state $v,j$ to state $v',j'$ are given in Ref.~\cite{selim:21}. Since
the total angular momentum $J$ and the overall parity $P$ are exact
quantum numbers, the CC equations can be solved separately for parities
$P = \pm 1$ and for all values of $J$ required to obtain converged cross
sections.

\subsection{Coupled states approximation}
\label{sec:CSA}
In the coupled-states approximation (CSA), introduced long ago
\cite{mcguire:74}, one neglects the Coriolis coupling terms off-diagonal
in $\Omega$ that appear in the second line of Eq.~(\ref{eq:T}). This
makes $\Omega$ an exact quantum number, so that the CC equations can be
separated into subsets of equations for each value of $\Omega$. The
dimension of each subset is smaller by a factor of $\min(2J+1,2j_{\rm
max}+1)$, where $j_{\rm max}$ is the maximum $j$-value in the basis,
than the dimension of the full CC equations. Since the CPU time to solve
the coupled equations is proportional to the third power of their
dimension, this yields a large reduction in computer time. Moreover,
since different $\Omega$ values are not coupled, the largest absolute
value of $\Omega$ is limited to the smallest of the initial or final $j$
value in the scattering process, which is smaller than $j_{\rm max}$ ---and
also much smalller than $J$ in most cases--- so the number of equation
subsets to be solved is small also.

In the standard application of the CSA the whole angular kinetic
operator in Eq.~(\ref{eq:coriolis}) is replaced by an operator
$\hat{L}^2$, with eigenvalues $L_{\rm eff}(L_{\rm eff}+1)$. The possible
values of $L$ range from $|J-j|$ to $J+j$, and different choices of
$L_{\rm eff}$ have been investigated. One mostly uses
$L_{\rm eff} = J$, which yields an angular kinetic energy proportional
to $J(J+1)$. In our BF implementation with the renormalized Numerov
propagator we have three different options. First, we can follow the
original CSA algorithm by using $L_{\rm eff} = J$ also in the matching
of the asymptotic scattering wave functions to obtain the scattering
matrices $\bmS^{P\,J\,\Omega}$ for all values of $P$, $J$, and $\Omega$.
The cross sections can then be obtained from the equation
\begin{equation}
\label{eq:ICS_CSA}
\sigma_{v',j' \leftarrow v,j}(E) = \frac{\pi}{(2{j}+1) k^{2}_{vj}}
   \sum_{P\,J\,\Omega} (2J+1) \left|
    \delta_{v'v} \delta_{j'j} -
  S^{P J \Omega}_{v'j';vj}(E)\right|^{2}.
\end{equation}

Instead of using each of the matrices $\bmQ_n^\Omega$ obtained after
solving the coupled-states equations for different $\Omega$ values
directly in an asymptotic matching procedure that yields matrices
$\bmS^{P\,J\,\Omega}$, one may follow an alternative procedure. This
alternative implies that the individual matrices $\bmQ_n^\Omega$ are
collected into a total matrix $\bmQ_n$, which consists of diagonal
subblocks with the matrices $\bmQ_n^\Omega$ for all $\Omega$ values. All
other elements of $\bmQ_n$ are zero, since in CSA there is no coupling
between functions with different $\Omega$. Since this total $\bmQ_n$
matrix involves the full BF channel basis containing all values of
$\Omega$, it can be transformed to its SF equivalent in the same way as
in the full CC treatment, see Eq.~(\ref{eq:transQ}). It can then be used
in the same asymptotic matching procedure as described in
Sec.~\ref{sec:CC} to obtain the scattering matrices $\bmS^{P\,J}$ that
occur in Eq.~(\ref{eq:ICS}) for the cross sections. In the propagation
of the individual matrices $\bmQ_i^\Omega$ with $i = 1, \ldots, n$ one
may either use the full diagonal angular kinetic energy $\big[J(J+1) +
j(j+1) - 2 \Omega^2\big]/2 \mu R^2$ from the first line of
Eq.~(\ref{eq:T}) or an effective angular kinetic energy $L_{\rm
eff}(L_{\rm eff}+1) = J(J+1)$ as in the original CSA method. Altogether,
this yields three different variants of the CSA, of which we compare the
results in Sec.~\ref{sec:results}. In the figures we label these
variants with $L_{\rm eff} = J$ for the standard CSA application, and
with the angular kinetic energies $\big[J(J+1) + j(j+1) - 2 \Omega^2\big]/2
\mu R^2$ and $J(J+1)$ that we take into account in the latter two CSA
methods.

\subsection{Improving the coupled-states approximation with first-order
Coriolis coupling}
\label{sec:NNCC}
The error in the CSA relative to full CC calculations is sometimes not
acceptable. Recently Yang \textit{et al.} \cite{yang:18} introduced an
improvement of the CSA method, called NNCC, which includes the
Coriolis coupling between functions with $\Omega$ and $\Omega \pm 1$ to
first order. It implies that the CC equations are still solved
separately for each $\Omega$ value, but with the off-diagonal Coriolis
couplings to the neighboring blocks with $\Omega \pm 1$ included in the
matrix $\bmW$, cf.\ Eqs.~(\ref{eq:Wmat}) and (\ref{eq:T}). In the full
CC equations these neighboring blocks are again coupled to functions
with $\Omega \pm 2$, but the latter are not directly coupled to the
functions with the given $\Omega$, so they only give rise to second and
higher order contributions to the solutions. The NNCC method
neglects these higher order Coriolis couplings.

The basis to solve the CC equations for each $\Omega$ in the NNCC
method involves also the bases for $\Omega-1$ and $\Omega+1$. The
propagation can then be done with the full angular kinetic energy from
Eq.~(\ref{eq:T}). Functions with different $\Omega$ are coupled by the
off-diagonal Coriolis coupling terms in the second line of this
equation, which has the effect that $\Omega$ is not a good quantum
number anymore. We could still label the separate propagations with
$\Omega$, but in order to distinguish the individual propagations from
the quantum numbers, we label them with the index $\oOmega$. The
propagation does not have to be performed for all $\Omega$ values,
because the $\Omega = 0$ basis is already included in the calculation
for the parity-adapted basis with $\tilde{\Omega} = 1$, see
Eq.~(\ref{eq:symbas}), and the $\tilde{\Omega} = J$ basis is included in
the calculation for $\tilde{\Omega} = J-1$. Also in the NNCC method
the calculations can be further restricted to $\tilde{\Omega}$ values
limited by the smallest of the initial and final $j$ quantum numbers,
augmented by one in this case. At the end of the propagations we have
two options. The first one is that we follow the NNCC method as
proposed by Yang \textit{et al.} \cite{yang:18}. At the end of each
propagation $\oOmega$ they construct an equivalent SF basis by
diagonalizing the operator $\hat{L}^2$ with the BF matrix elements from
Eq.~(\ref{eq:T}) in the space of functions $\Omega-1$, $\Omega$, and
$\Omega+1$. The eigenvalues $L(L+1)$ from this diagonalization are
non-integer, and also $L$ is therefore non-integer. The same procedure
has been implemented previously in the reactive scattering program ABC
\cite{skouteris:00}, which has the possibility to truncate the
$\Omega$ basis.
The matrix $\bmB^{\oOmega}$ with the eigenvectors of $\hat{L}^2$ in the
restricted $\Omega$ space as column vectors is then used to transform
the matrix $\bmQ_n^{\oOmega}$ obtained at the end of propagation
$\oOmega$ to its SF equivalent
\begin{equation}
\label{eq:transQNN}
  \bmQ^{\oOmega,\textrm{SF}}_n =
       {\bmB^{\oOmega}}^{\,\dagger} \bmQ_n^{\oOmega} \bmB^{\oOmega}.
\end{equation}
This numerical diagonalization of $\hat{L}^2$ leaves the overall sign of
each eigenvector, i.e., of each column of the matrix $\bmB^{\oOmega}$,
undetermined. This has no effect on the final results, however, because
the inverse of $\bmB^{\oOmega}$ is used in Eq.~(\ref{eq:transback}) to
transform the $S$-matrix back from the SF to the BF frame. Indeed, we
found numerically that the final cross sections in Eq.~(\ref{eq:NNCC})
do not depend on these signs. Since the SF basis thus obtained
corresponds to non-integer $L$ values, the spherical Bessel functions
used in the matching procedure described in
Eqs.~(\ref{eq:psi_asymptotic}) to (\ref{eq:matK}) must be the
corresponding functions with non-integer $L$ values
\cite{abramowitz:64}.

The second option is that we consider the matrices $\bmQ^{\oOmega}_n$
over the BF basis with $\Omega-1$, $\Omega$, and $\Omega+1$ as part of
the matrix $\bmQ_n$ over the full BF basis and transform them to the SF
basis with the aid of Eq.~(\ref{eq:transQ}). The transformation matrix
$\bmU$ contains Clebsch-Gordan coefficients, the $L$ values are
integers, and the asymptotic matching procedure to obtain SF matrices
$\bmS^{{PJ\oOmega (\textrm{SF}})}$ from each propagation $\oOmega$ can
be done with the usual spherical Bessel functions.

Finally, in both options, one has to transform the matrices
$\bmS^{PJ\oOmega\,(\textrm{SF})}$ in the SF basis back to the BF
basis with the aid of the inverse BF to SF transformation. In the first
option, this transformation is done with the matrix $\bmB^{\oOmega}$ for the given
$\oOmega$. Since $\bmB^{\oOmega}$ is a real orthogonal matrix, its inverse is
simply its transpose and one obtains
\begin{equation}
\label{eq:transback}
S^{PJ\oOmega(\textrm{BF})}_{v'j'\Omega';vj\Omega} = \sum_{L'L}
i^{L-L'} B^{\oOmega}_{\Omega' L'}
S^{PJ\oOmega (\textrm{SF})}_{v'j'L';vjL} B^{\oOmega}_{L \Omega}
\end{equation}
In the second option, the BF to SF transformation is done with the
matrix $\bmU$ from Eq.~(\ref{eq:transQ}), which is also orthogonal, and
one obtains the same formula, with $\bmB^{\oOmega}$ replaced by $\bmU$. In order
to avoid double counting, we select from each matrix
$\bmS^{PJ\oOmega\,\textrm{BF}}$ only the blocks with $\Omega=\oOmega$
and calculate the cross sections with the formula
\begin{equation}
\label{eq:NNCC}
\sigma_{v',j' \leftarrow v,j}(E) = \frac{\pi}{(2{j}+1) k^{2}_{vj}}
   \sum_{\Omega} \sum_{\Omega'=\Omega-1}^{\Omega+1}
   \sum_{P\,J\,\Omega} (2J+1) \left|
    \delta_{v'v} \delta_{j'j} -
  S^{PJ\oOmega (\textrm{BF})}_{v'j'\Omega';vj\Omega}(E)\right|^{2}.
\end{equation}

\subsection{Combining NNCC with the multi-channel distorted wave Born
approximation}
\label{sec:DWBA}
In a previous paper \cite{selim:21} we have shown how the application of
a multi-channel distorted wave Born approximation (MC-DWBA) in
scattering calculations reduces the computer time by about a factor of
three with respect to exact CC calculations, but produces cross sections
and rate coefficients for rovibrational transitions in CO$_2$-He
collisions that are about equally accurate. The MC-DWBA algorithm and
the reason why it can be favorably applied to rovibrational transitions
are explained in detail in Ref.~\cite{selim:21}. Here we investigate the
application of this algorithm in combination with the NNCC method.

\subsection{Computational details}
\label{sec:detail}
We study state-to-state ($v,j \rightarrow v',j'$) rovibrational
transitions in CO$_2$ by collisions with He in which CO$_2$ is
de-excited from the $v=1$ symmetric stretch fundamental to the $v'=0$
ground state. The details of the calculation and the characteristics of
the 3D intermolecular potential of CO$_2$-He with CO$_2$ deformed along
the symmetric stretch coordinate are given in
Ref.~\onlinecite{selim:21}. The basis in the scattering calculations
consisted of CO$_2$ symmetric stretch functions with $v = 0$ and 1, the
rotational basis contained all functions with $j \le 70$, for both $v =
0$ and 1. Calculations were made with a larger vibrational basis
including also $v=2$ functions, but this made practically no difference
for the results. It is interesting that the states with $v,j = 1,0$ and
$v,j = 0,58$ have nearly the same energy, so that transitions between
these states are nearly resonant. The radial grid for the renormalized
Numerov propagator contained 502 equidistant points in the range $3 \le
R \le 30~a_0$.
Rovibrational state-to-state cross sections were calculated for
collision energies up to 3000~cm$^{-1}$, with steps of 0.1~cm$^{-1}$ in
the resonance regime from $1\le E \le 20$~cm$^{-1}$, steps of
1~cm$^{-1}$ for $20\le E \le 50$~cm$^{-1}$, 2~cm$^{-1}$ for $50\le E \le
100$~cm$^{-1}$, 50~cm$^{-1}$ for $100\le E \le 1000$~cm$^{-1}$, and
200~cm$^{-1}$ for $1000\le E \le 3000$~cm$^{-1}$. The largest total $J$
value included was 100, for both parities $P = \pm 1$. The corresponding
rate coefficients were calculated for temperatures from 10 to 500\,K by
cubic spline interpolating the cross sections over the energy grid and
calculating the integral in Eq.~(14) of Ref.~\onlinecite{selim:21} with
the trapezoidal rule. Since the rotational states of CO$_2$ up to $j=50$
with energy 991~cm$^{-1}$ are populated at the highest temperature, we
calculated the cross sections and rate coefficients for initial states
up to this value of $j$.

\begin{figure}[ht!]
\includegraphics*[width=0.79\textwidth]{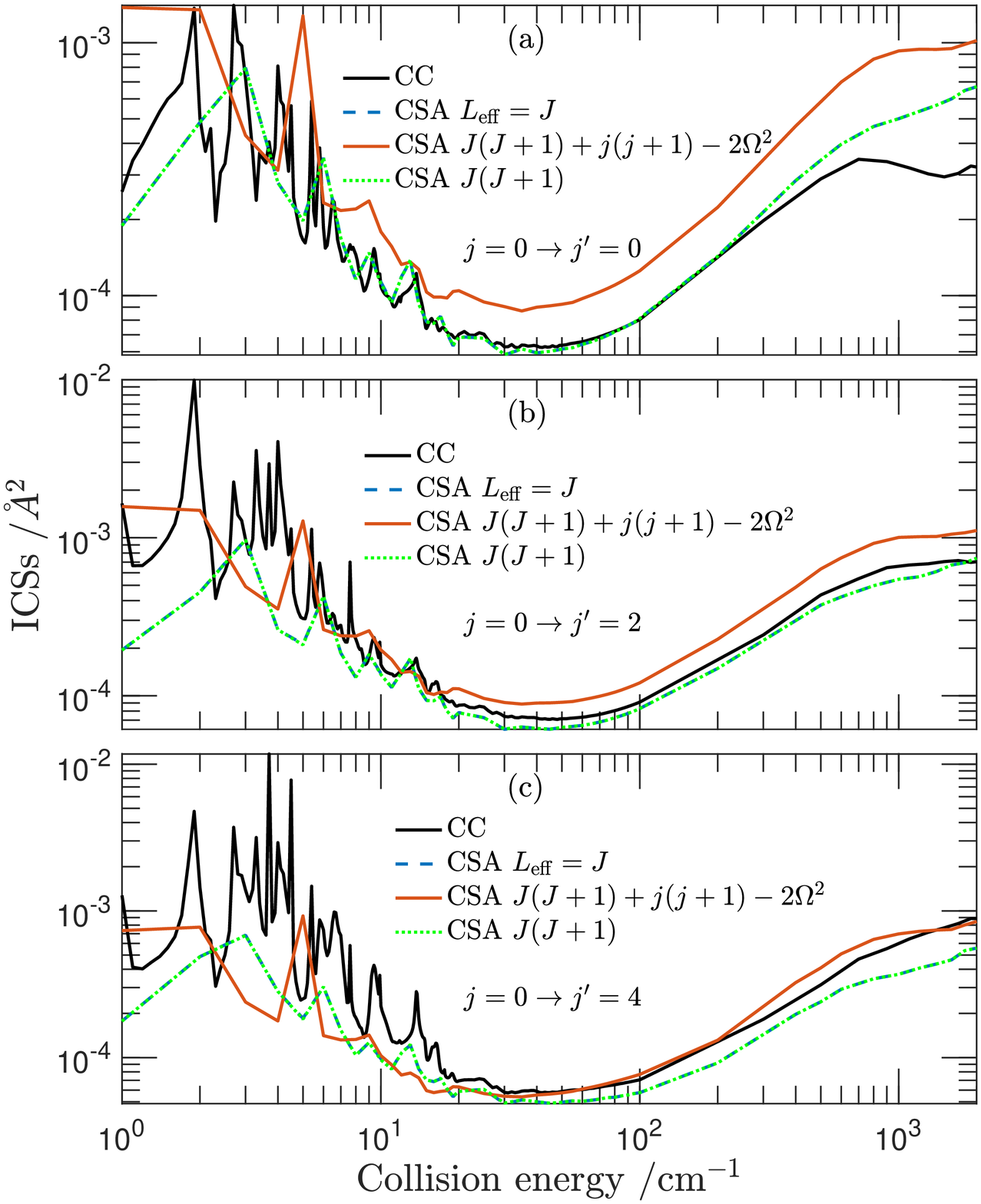}
\caption{ICSs from the three different CSA methods described in
Sec.~\ref{sec:CSA} compared with full CC results. The upper legend CC
refers to ICSs from full CC calculations. The next legend CSA $L_{\rm
eff}=J$ refers to the conventional CSA method with angular kinetic
energy $L_{\rm eff}(L_{\rm eff}+1)$ and $L_{\rm eff} = J$ in which the
ICS is calculated directly from the S-matrices for different $\Omega$ in
the BF frame. The lower two legends refer to the CSA methods with
angular
kinetic energy terms $J(J+1) + j(j+1) - 2 \Omega^2$ or $J(J+1)$ used in
the BF propagation and the ICS obtained from the $S$-matrix in the SF
frame, after transformation of the BF matrix $\bmQ_n$ to the SF frame
with the aid of Eq.~(\ref{eq:transQ}). The initial state is $v=1,j=0$,
the final states are $v'=0$ with $j'=0$ (a), $j'=2$ (b), and $j'=4$ (c).}
\label{fig:CSA1}
\end{figure}
\begin{figure}[ht!]
\includegraphics*[width=0.79\textwidth]{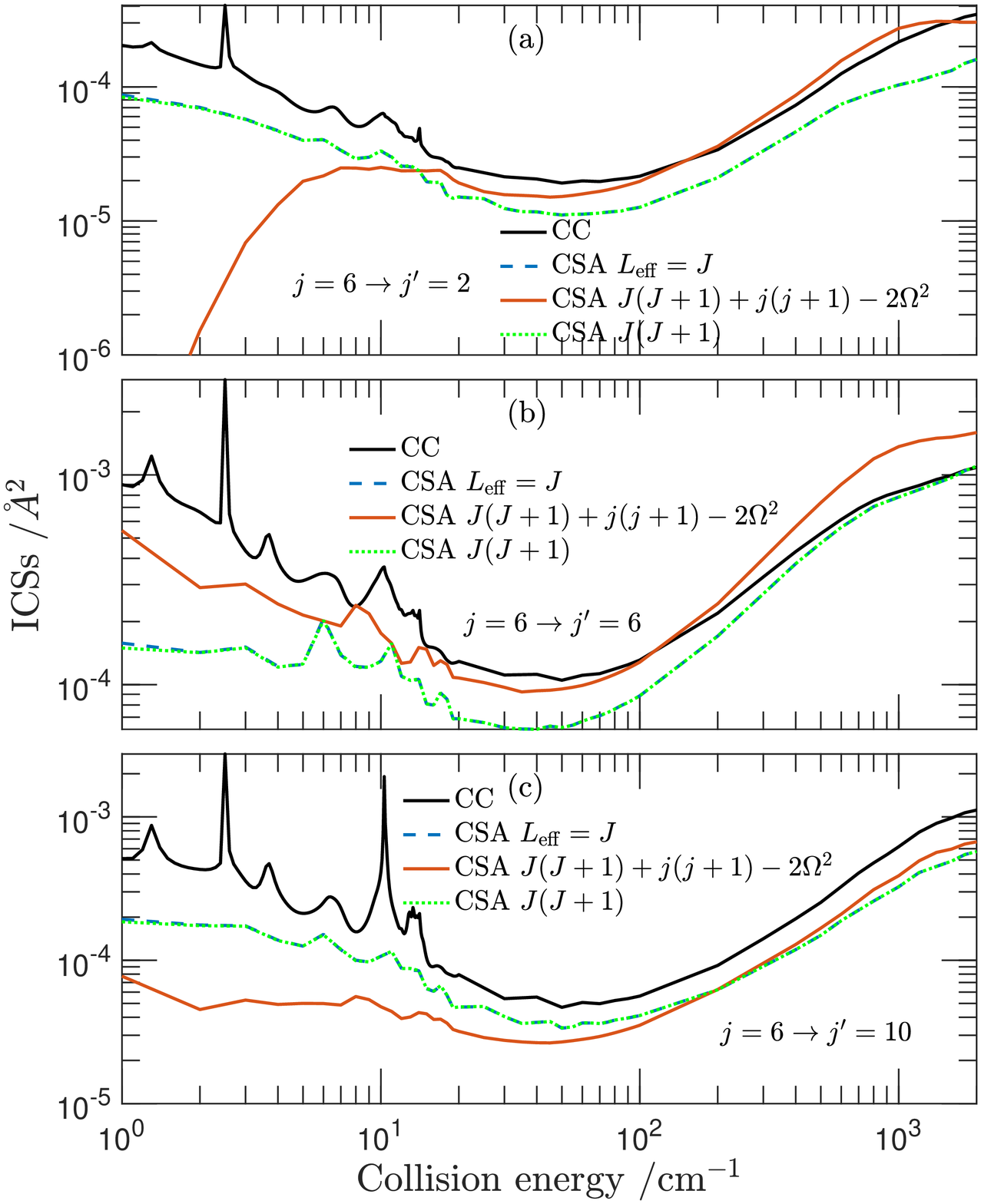}
\caption{Same as Fig.~\ref{fig:CSA1}, with a different initial state:
$v=1,j=6$ and final states $v'=0$ with $j'=2$ (a), $j'=6$ (b), and
$j'=10$ (c).}
\label{fig:CSA2}
\end{figure}

\section{Results and discussion}
\label{sec:results}
Figures~\ref{fig:CSA1} and \ref{fig:CSA2} show the integral cross
sections (ICSs) from the three different CSA methods described in
Sec.~\ref{sec:CSA}, in comparison with ICSs from full CC calculations,
for quenching from initial state $v,j=1,0$ to different final states
$v'=0,j'$. The meaning of the legends referring to the different CSA
methods is explained in the caption of the figure. The sharp peaks in
the ICSs for collision energies below 20~cm$^{-1}$ correspond to
resonances, which are extremely sensitive to the details of the
calculations. Hence, it is not surprising that they are different for
the different scattering methods used. But also for higher energies one
observes that the ICSs from the CSA methods differ substantially from
the full CC results. Two of the CSA methods produce very similar
results. In both of them the propagation is done in the BF frame with
the angular kinetic energy term $J(J+1)$, but the asymptotic matching
procedures to obtain the $S$-matrix are different, see
Sec.~\ref{sec:CSA}. Apparently, the way of matching is less important
than the approximation in the angular kinetic energy term used in the
propagation. The third CSA method, with the angular kinetic energy term
$J(J+1) + j(j+1) - 2 \Omega^2$ produces rather different ICSs. The
deviations of the CSA ICSs from the CC results are rather erratic, and
one cannot conclude that one of the CSA methods is definitely better
than the others. The method with the term $J(J+1) + j(j+1) - 2 \Omega^2$
contains the full angular kinetic energy when neglecting the Coriolis
coupling between basis functions with different $\Omega$, so in the
following we will show the results from this particular CSA
approximation.
\begin{figure}[ht!]
\includegraphics*[width=0.89\textwidth]{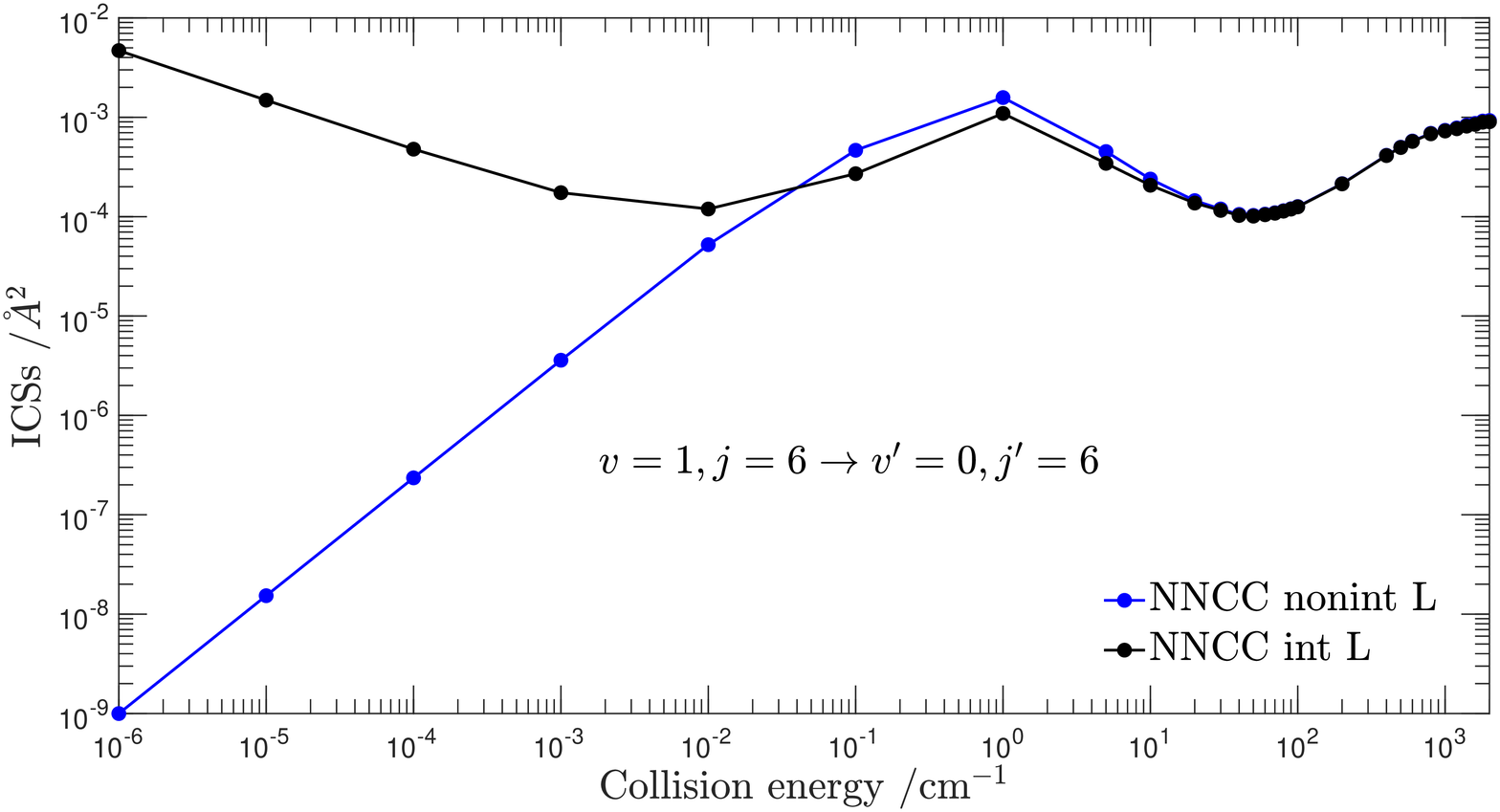}
\caption{Comparison of ICSs calculated with the two different NNCC
methods described in Sec.~\ref{sec:NNCC}. They differ in the asymptotic
matching procedure using integer $L$ or non-integer $L$ values.}
\label{fig:match}
\end{figure}

Figure~\ref{fig:match} illustrates the effect of the different
asymptotic matching procedures used in the NNCC methods described in
Sec.~\ref{sec:NNCC}. In the method proposed by Yang \textit{et al.}
\cite{yang:18} they numerically determine the eigenvectors of the
operator $\hat{L}^2$ in the BF basis restricted to a given $\Omega$ and
$\Omega \pm 1$, which yields non-integer eigenvalues and, hence,
non-integer $L$ values. These non-integer $L$ values are then used in
the asymptotic matching procedure that yields the $S$-matrix. In the
second method, we use specific rows of the matrix $\bmU$ from
Eq.~(\ref{eq:U}) that transforms the full BF basis to the SF basis. This
full BF basis corresponds to a SF basis with integer $L$ values, which
are used in the asymptotic matching procedure. Figure~\ref{fig:match}
shows that the ICSs from the two methods are very similar for collision
energies higher than about 10~cm$^{-1}$. For lower energies, between 0.1
and 10~cm$^{-1}$, they become different, but this is the resonance
regime where the ICSs are very sensitive to details of the potential and
the computational method. The difference becomes very large for still
lower energies approaching the Wigner regime. In this regime the ICSs of
inelastic collisions should depend on $E^{-1/2}$ according to the Wigner
threshold laws \cite{wigner:48}. The method with integer $L$ values
nicely obeys this relation, but the method with non-integer $L$ values
fails completely. This is not surprising, because the Wigner threshold
law \cite{wigner:48} assumes pure $s$-wave scattering at the limit of
very low energies, which implies that $L=0$. Only the method with
integer $L$ correctly reaches this limit.
\begin{figure}[ht!]
\includegraphics*[width=0.49\textwidth]{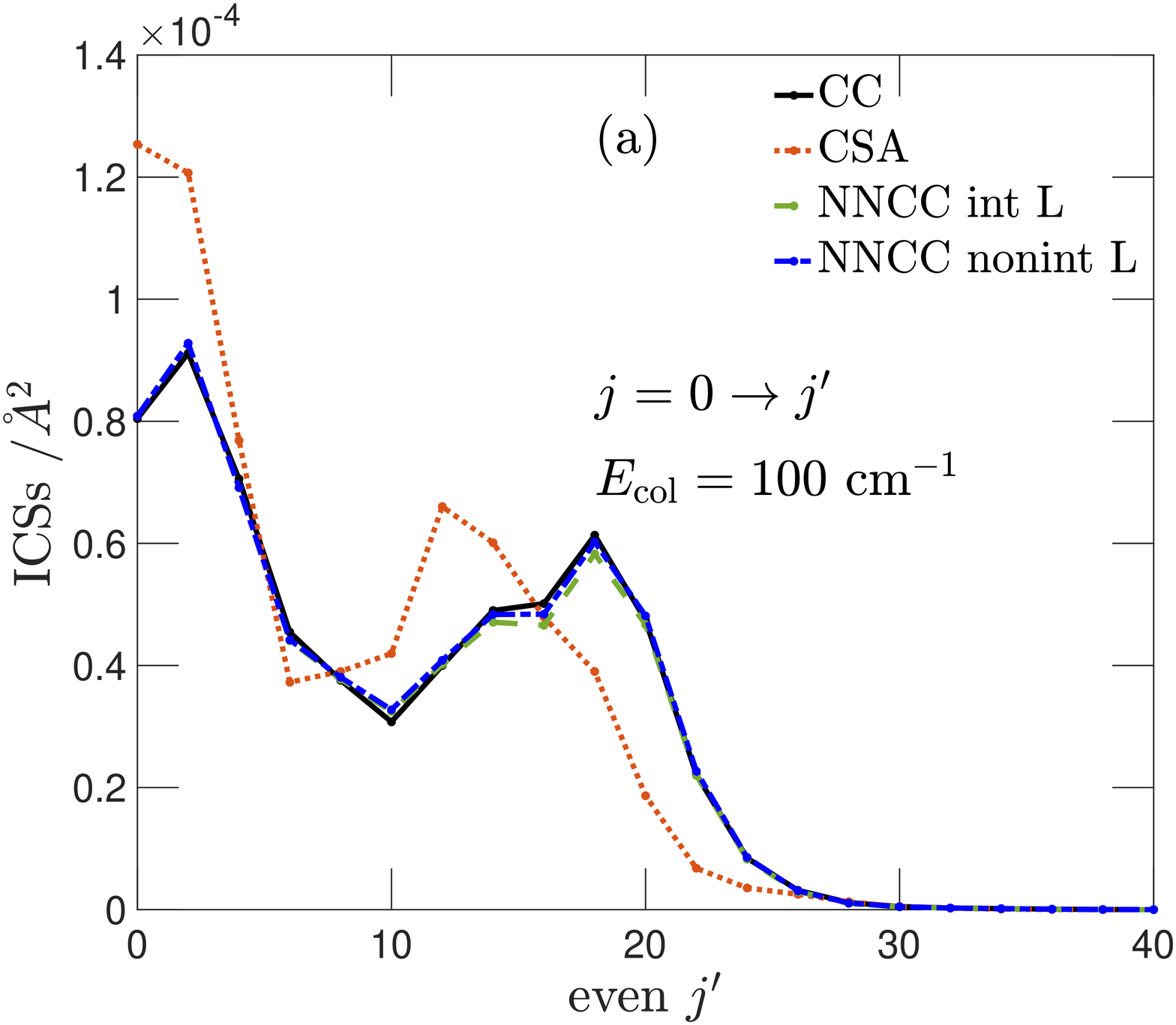}
\includegraphics*[width=0.48\textwidth]{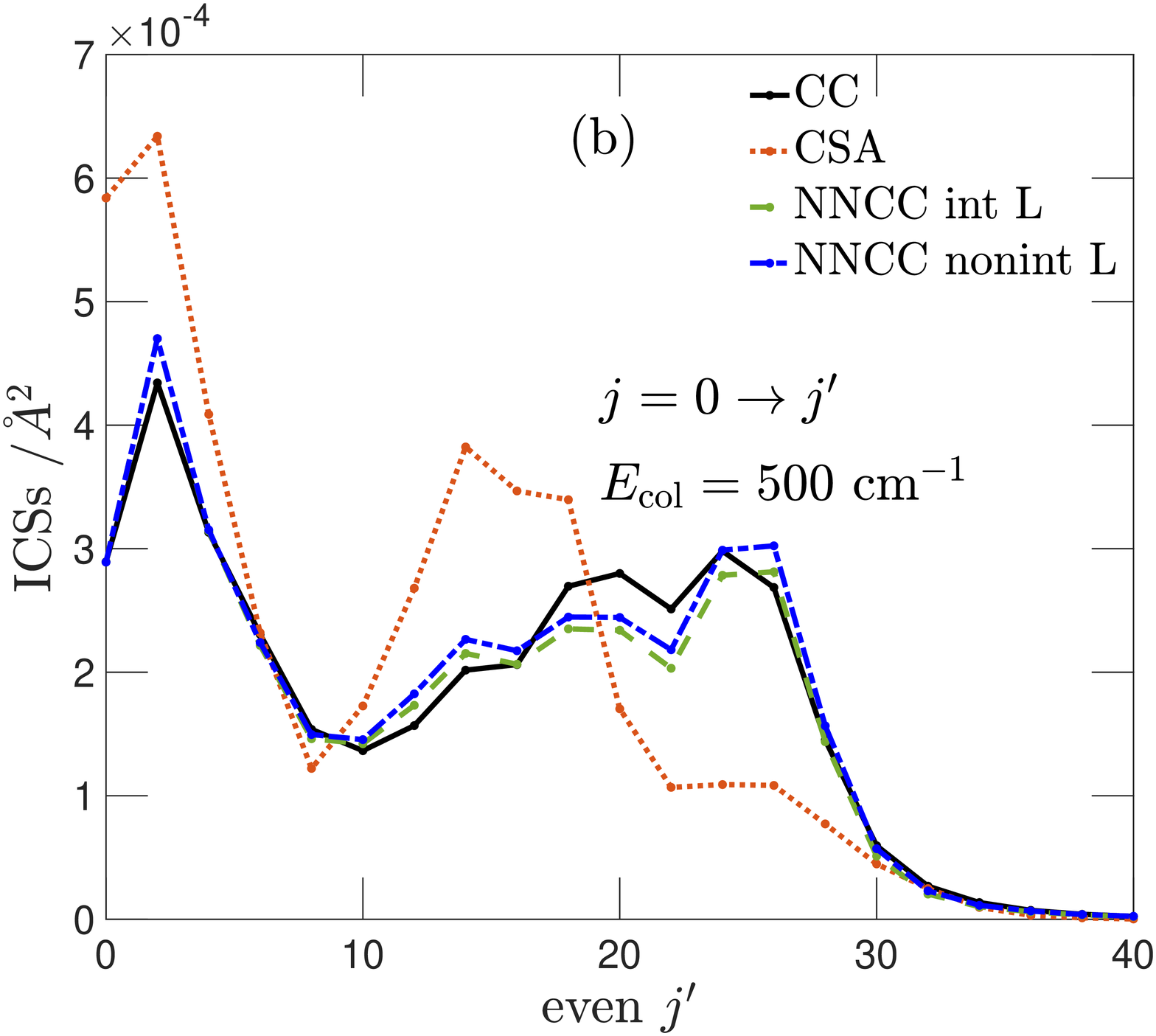}
\caption{Product ($v'=0,j'$) distributions from the NNCC methods with
integer and non-integer $L$ values for initial state $v=1,j=0$, compared
with CSA and full CC calculations, at collision energies $E =
100$~cm$^{-1}$ (a) and 500~cm$^{-1}$ (b). Note that only even values of
$j$ and $j'$ are physically allowed. The CSA method contains the angular
kinetic energy $J(J+1) + j(j+1) - 2 \Omega^2$.}
\label{fig:inj0}
\end{figure}
\begin{figure}[ht!]
\includegraphics*[width=0.49\textwidth]{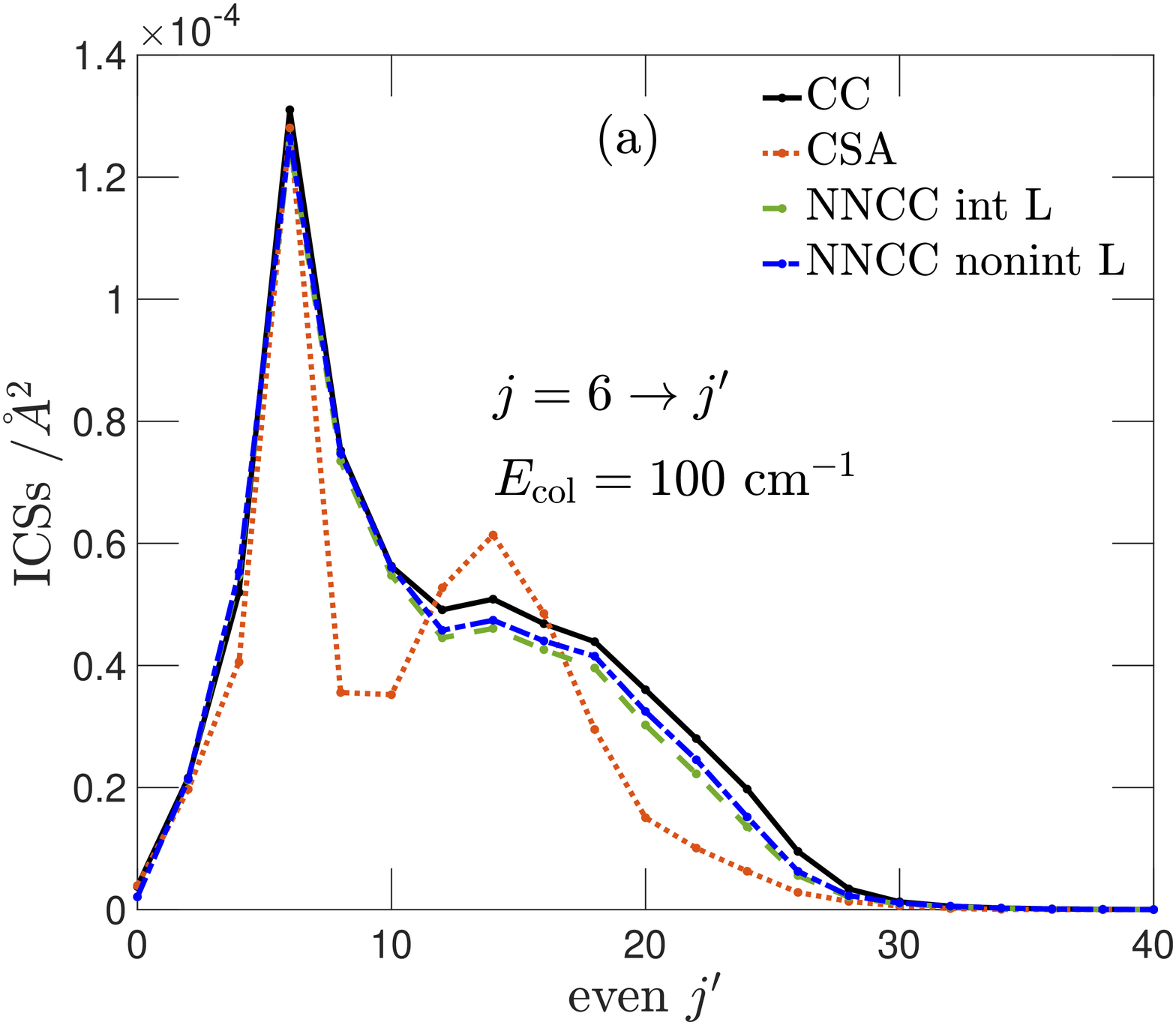}
\includegraphics*[width=0.48\textwidth]{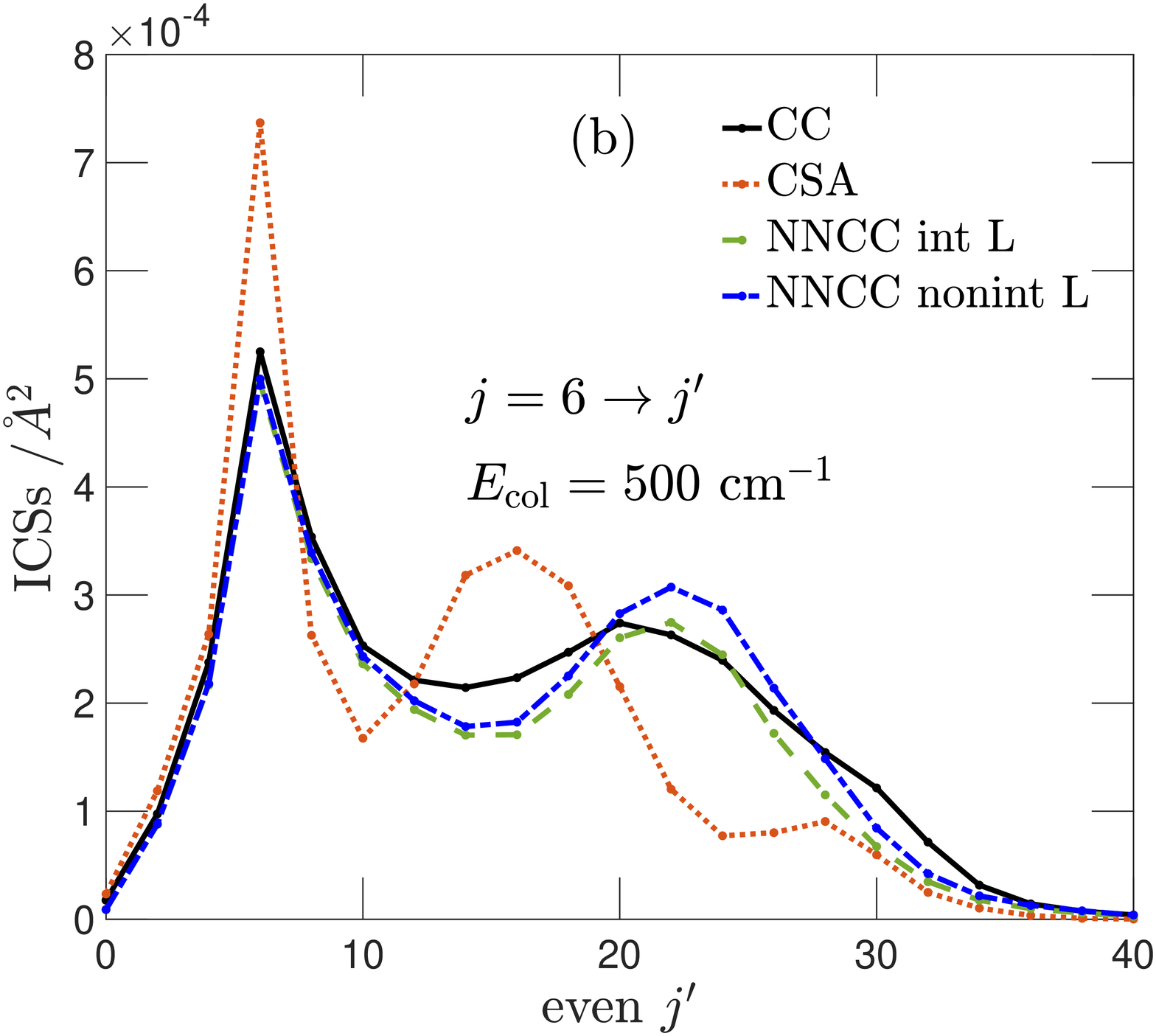}
\caption{Same as Fig.~\ref{fig:inj0} for initial state $v=1,j=6$ at
collision energies $E = 100$~cm$^{-1}$ (a) and 500~cm$^{-1}$ (b).}
\label{fig:inj6}
\end{figure}
\begin{figure}[ht!]
\includegraphics*[width=0.79\textwidth]{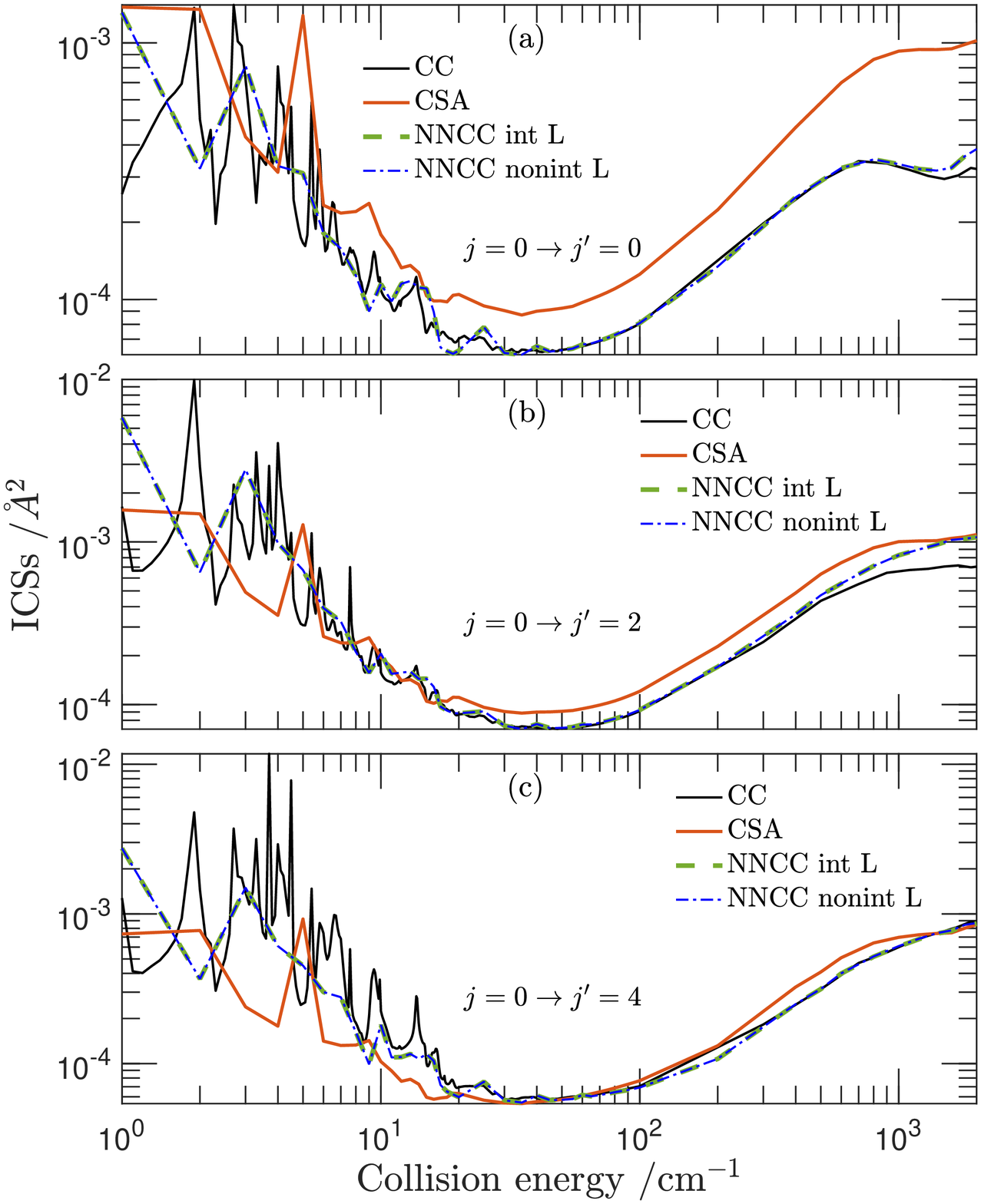}
\caption{ICSs from the NNCC methods with integer and non-integer $L$
values described in Sec.~\ref{sec:NNCC} compared with CSA and full CC
results. The ICSs are shown as functions of the collision energy $E$ for
initial state $v=1,j=0$ and final states $v'=0,j'=0$ (a),
$v'=0,j'=2$ (b), and $v'=0,j'=4$ (c). The CSA method contains the
angular kinetic energy $J(J+1)+ j(j+1) - 2 \Omega^2$.}
\label{fig:NNCC1}
\end{figure}
\begin{figure}[ht!]
\includegraphics*[width=0.79\textwidth]{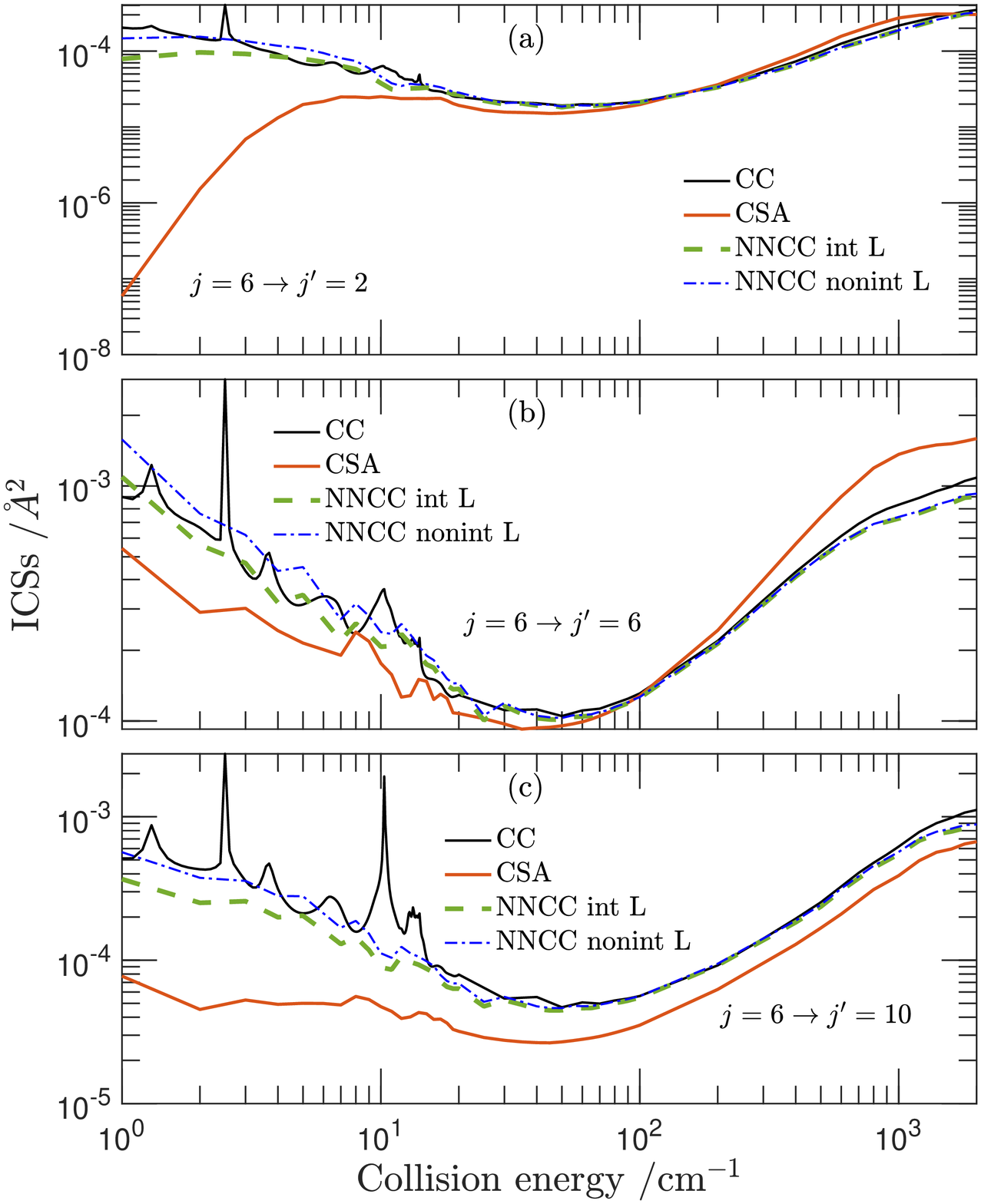}
\caption{Same as Fig.~\ref{fig:NNCC1} for initial state $v=1,j=6$ and
final states $v'=0,j'=2$ (a), $v'=0,j'=6$ (b), and $v'=0,j'=10$ (c).}
\label{fig:NNCC2}
\end{figure}

Figures~\ref{fig:inj0} and \ref{fig:inj6} show product distributions
from the NNCC methods with integer and non-integer $L$ values as
described in Sec.~\ref{sec:NNCC}, for different initial states and
different collision energies, compared with results from CSA and full CC
calculations. A similar comparison is made in Figs.~\ref{fig:NNCC1} and
\ref{fig:NNCC2}, which show the ICSs for the same initial $v=1,j$ values
and specific final $v'=0,j'$ values as functions of the collision
energy. It is obvious from all these figures that the NNCC method, which
includes the first-order Coriolis coupling between basis functions with
$\Omega$ and $\Omega \pm 1$, produces results in much better agreement
with full CC calculations than the CSA methods. The NNCC methods with
integer and non-integer $L$ values perform quite similarly.
Except, of course, for collision energies below 0.1~cm$^{-1}$ displayed
in Fig.~\ref{fig:match}, where the NNCC method with integer $L$ values
becomes better because it obeys the Wigner threshold law, as discussed
above.

\begin{figure}[ht!]
\includegraphics*[width=0.79\textwidth]{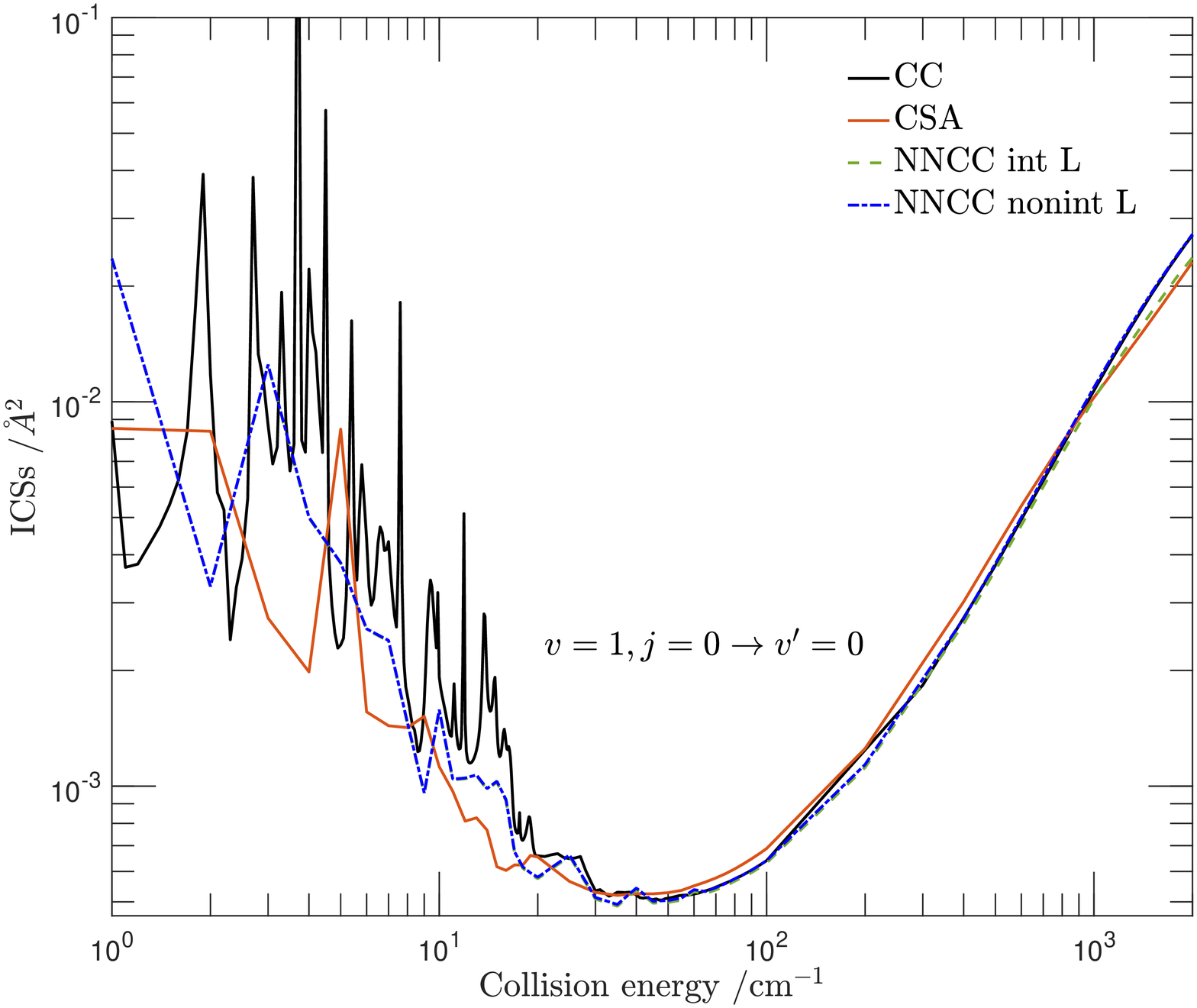}
\caption{ICSs for quenching from initial state $v=1,j=0$ to the $v'=0$
state, summed over all final $j'$ values. The different curves are
results from the NNCC methods with integer and non-integer $L$
values, compared with CSA and full CC results. The CSA method contains the
angular kinetic energy $J(J+1) + j(j+1) - 2 \Omega^2$.}
\label{fig:quench1}
\end{figure}

\begin{figure}[ht!]
\includegraphics*[width=0.79\textwidth]{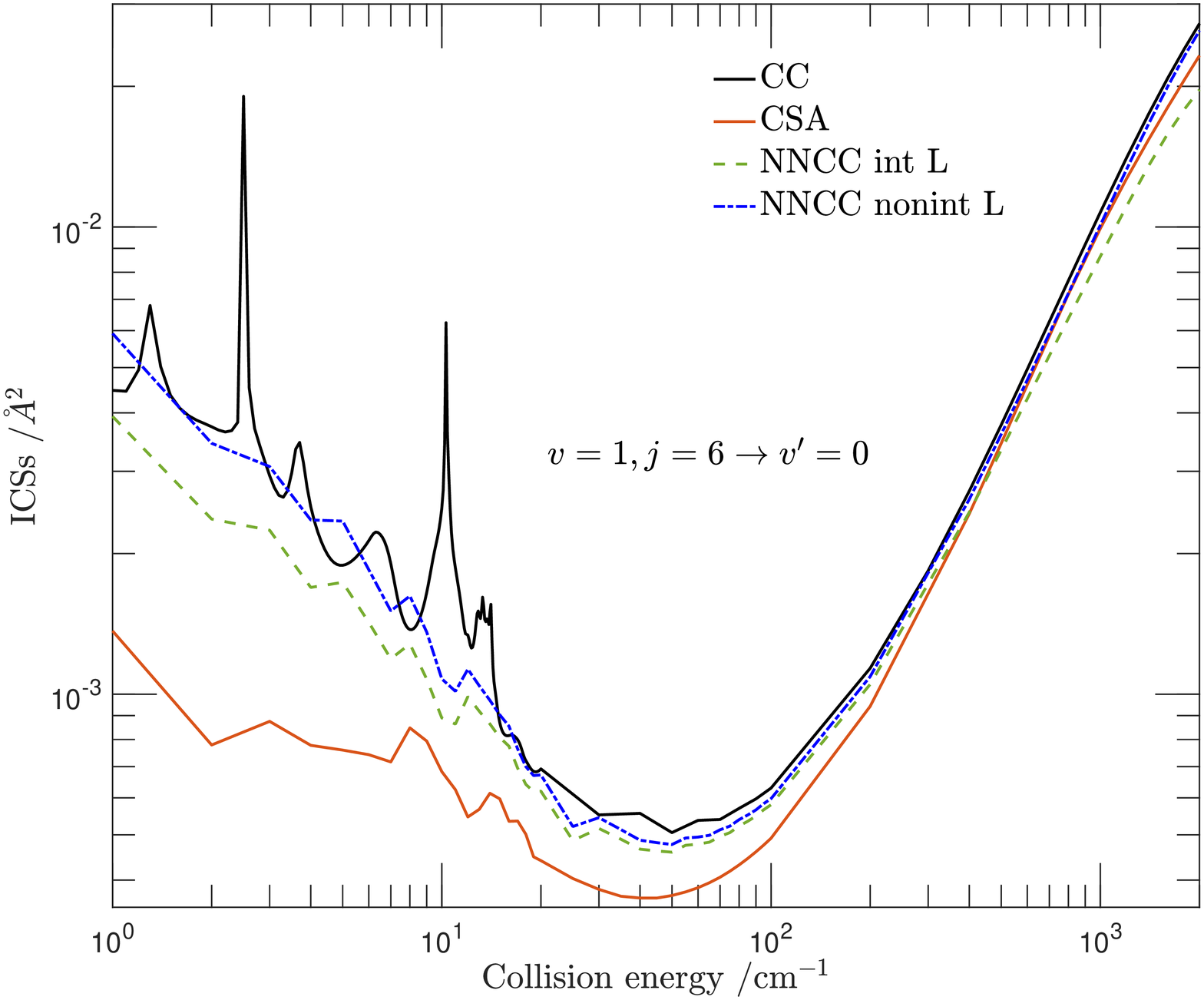}
\caption{Same as Fig.~\ref{fig:quench1} for initial state $v=1,j=6$.}
\label{fig:quench2}
\end{figure}

In Figs.~\ref{fig:quench1} and \ref{fig:quench2} we display the total
quenching cross sections for transitions from initial states $v=1,j=0$
and $v=1,j=6$ to all final $v'=0,j'$ states. The resonance structure in
the ICSs is clearly more pronounced for the $v=1,j=0$ initial state than
for the $v=1,j=6$ state. These resonances are so sensitive to the method
of computation that none of the methods can reproduce the CC results.
For energies higher than 20~cm$^{-1}$ above the resonance regime both
NNCC methods produce results in good agreement with full CC results for
both initial states $v=1,j=0$ and $v=1,j=6$. It seems somewhat
surprising that also the CSA method yields fairly good results for the
$v=1,j=0$ initial state (Fig.~\ref{fig:quench1}), but as one can see in
Fig.~\ref{fig:inj0} CSA considerably overestimates the ICSs for some of
the final $j'$ states and underestimates them for other $j'$ values, and
these errors nearly compensate each other. For the initial
$v=1,j=6$ state (Fig.~\ref{fig:quench2}) the CSA results are clearly
inferior to the ICSs from the NNCC methods for energies below
200~cm$^{-1}$.

A remarkable observation when comparing Figs.~\ref{fig:quench1} and
\ref{fig:quench2} is that the total quenching cross sections are almost
the same for the $v=1,j=0$ and $v=1,j=6$ initial states. We also
considered other initial $j$ values and we found, as in our studies in
Ref.~\cite{selim:21}, that these total quenching cross sections hardly
depend on the initial $j$ value. This is important as it implies that
one can apply our results for some specific initial $j$ values more
generally, to all different initial $j$ states and thus obtain total
$v=1 \rightarrow v'=0$ quenching cross sections and rate coefficients as
functions of the temperature.

\begin{figure}[ht!]
\includegraphics*[width=0.79\textwidth]{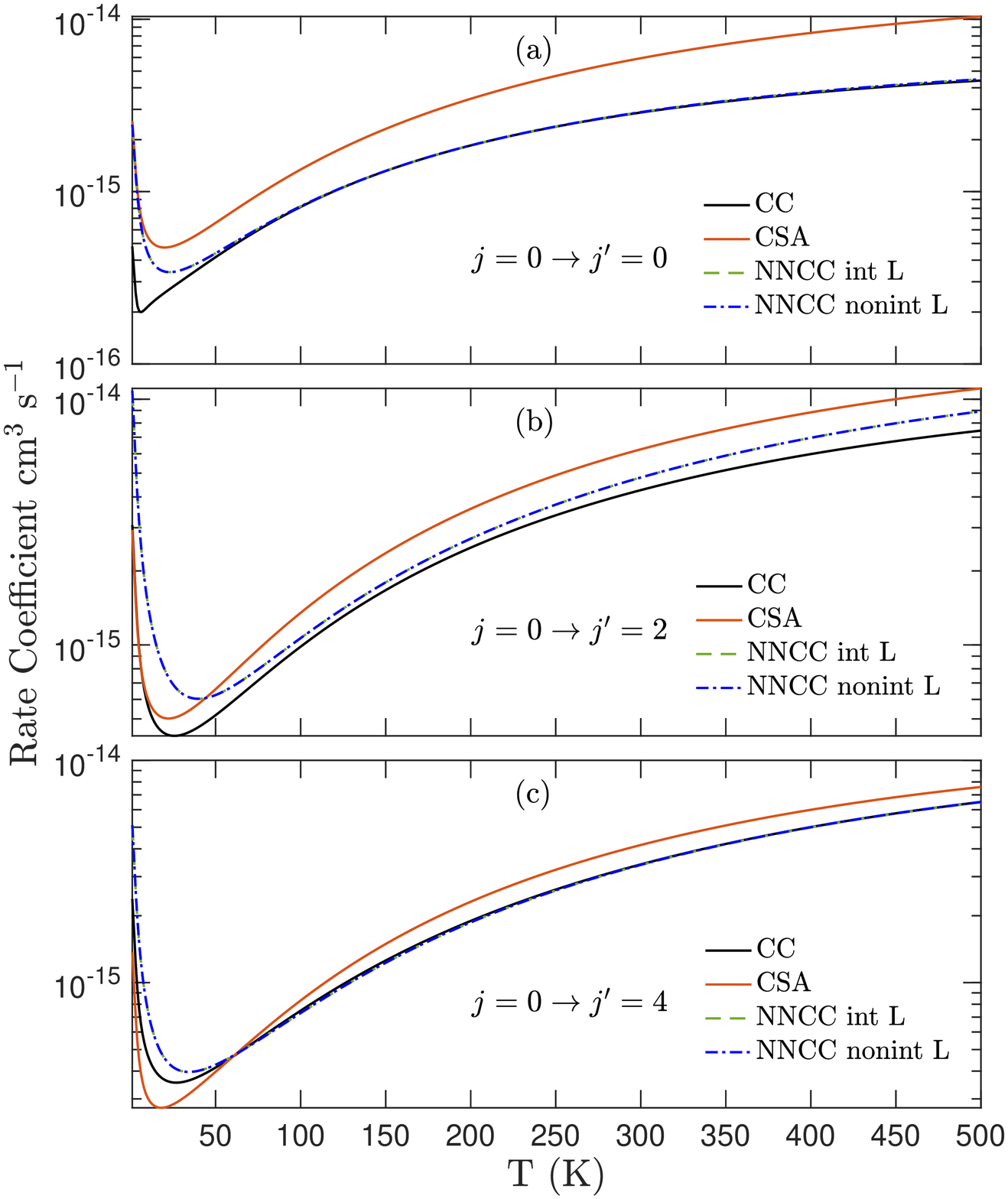}
\caption{State-to-state rate coefficients as functions of the temperature
calculated for initial state $v=1,j=0$ and final states $v'=0,j'=0$ (a),
$v'=0,j'=2$ (b), and $v'=0,j'=4$ (c), with the ICSs shown
in Fig.~\ref{fig:NNCC1} (a), (b), and (c).}
\label{fig:rates1}
\end{figure}

\begin{figure}[ht!]
\includegraphics*[width=0.79\textwidth]{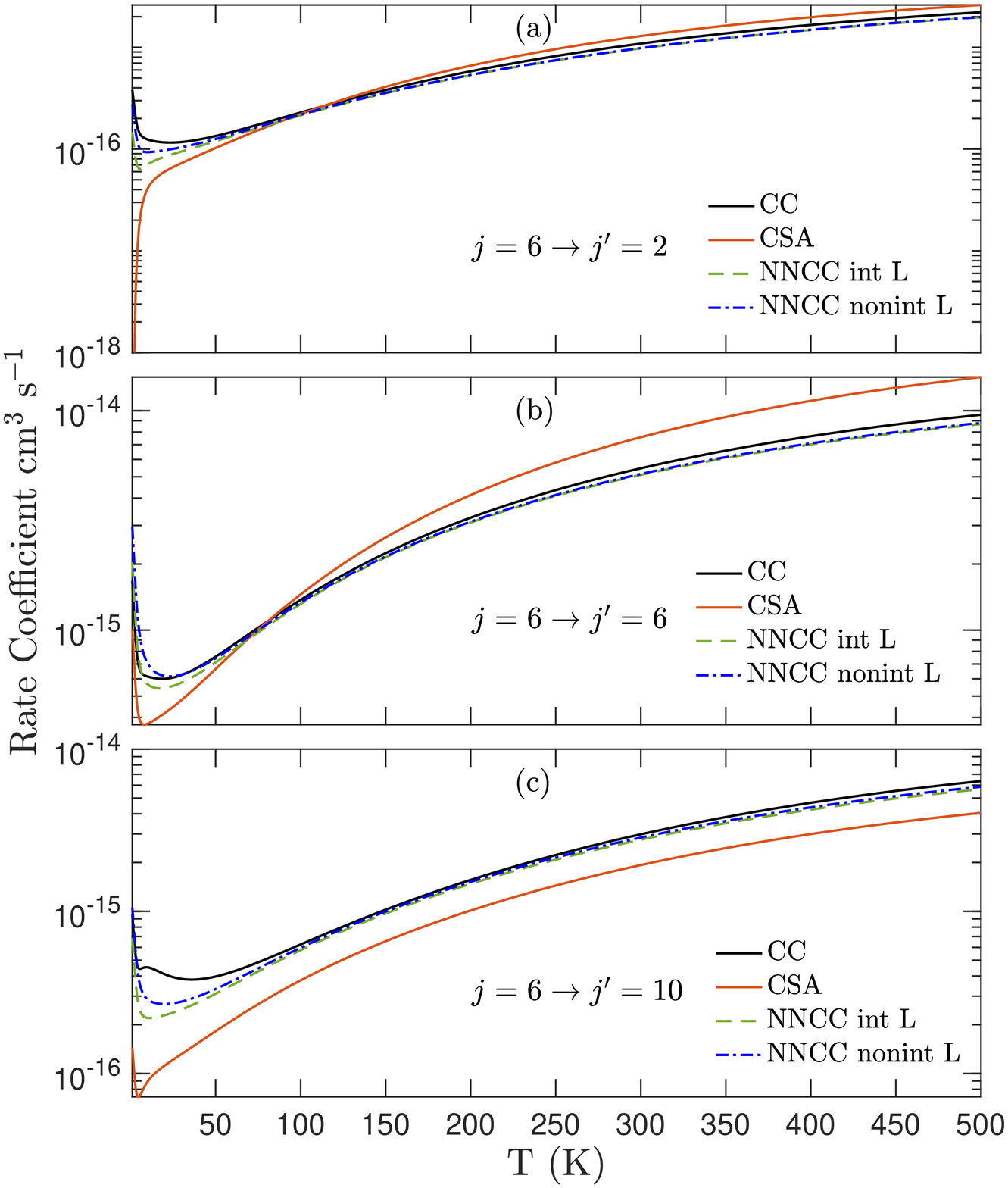}
\caption{State-to-state rate coefficients as functions of the temperature
calculated for initial state $v=1,j=6$ and final states $v'=0,j'=2$ (a),
$v'=0,j'=6$ (b), and $v'=0,j'=10$ (c), with the ICSs shown
in Fig.~\ref{fig:NNCC2} (a), (b), and (c).}
\label{fig:rates2}
\end{figure}

So far we discussed the ICSs from different methods as functions of the
collision energy, but in astrophysical modeling one needs collisional
rate coefficients as functions of the temperature.
Figures~\ref{fig:rates1} and \ref{fig:rates2} show state-to-state rate
coefficients calculated from the ICSs displayed in Figs.~\ref{fig:NNCC1}
and \ref{fig:NNCC2}. It is clear that the conclusions discussed above
for the ICSs apply also to the rate coefficients. That is, the NNCC
method is clearly superior to the CSA method in reproducing the full CC
results.

\begin{figure}[ht!]
\includegraphics*[width=0.95\textwidth]{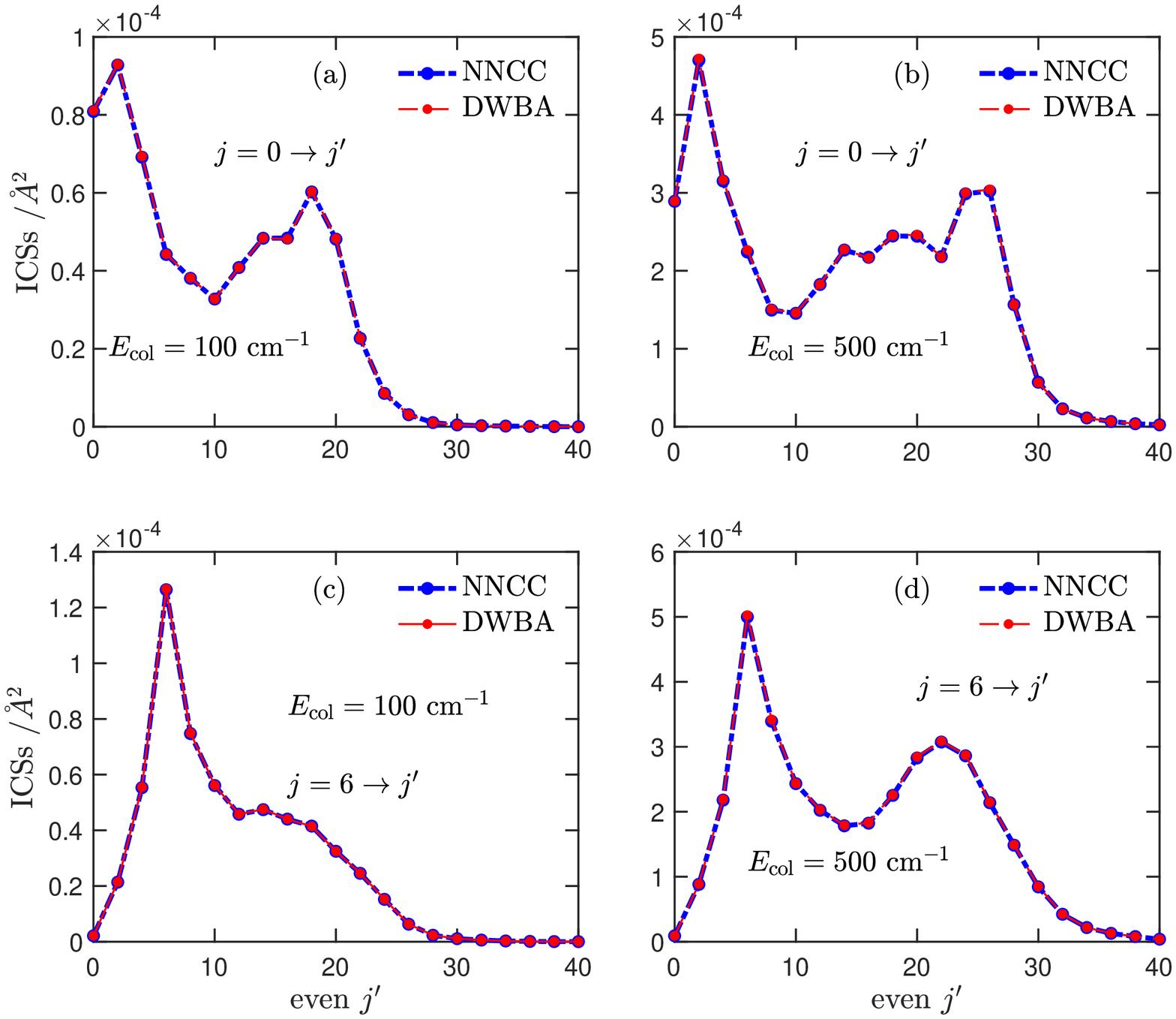}

\caption{Product ($v'=0,j'$) distributions from the NNCC method with
non-integer $L$ values, compared with results from MC-DWBA applied to
this method. Panels (a) and (b) are for initial state $v=1,j=0$ at
collision energies $E = 100$ and 500~cm$^{-1}$, respectively, panels (c)
and (d) for initial state $v=1,j=6$ at the same collision energies.}

\label{fig:prodDWBA}
\end{figure}

Finally, Fig.~\ref{fig:prodDWBA} demonstrates that state-to-state ICSs
for rovibrationally inelastic collisions from the NNCC method are
accurately reproduced by combining this method with MC-DWBA. This is
very useful, as it shows that MC-DWBA cannot only be applied to the
full CC method, as in Ref.~\cite{selim:21}, but also to more
approximate and less expensive methods.

Another issue to be discussed is: how much more efficient are the CSA
and NNCC methods than the full CC method, both in terms of computer
memory and CPU time. We checked this in calculations with a basis of
CO$_2$ rotational functions up to $j = 100$ for $v = 0$ and 1, and total
angular momentum $J = 100$. As initial state we took $v=1,j=6$, so in
the CSA calculations we needed to solve seven sets of coupled-channel
equations for $\tilde{\Omega}$ in the symmetry-adapted basis ranging
from 0 to 6. The NNCC method includes blocks with $\Omega$ and $\Omega
\pm 1$ and we had to make calculations for $\tilde{\Omega}$ = [0,1,2],
[1,2,3], [2,3,4], [3,4,5], [4,5,6], and [5,6,7]. The number of channels
in the CC calculations was 5202, in CSA it was 100, and in NNCC it was
290. The size of the matrices involved in solving the coupled-channel
problem depends quadratically on the number of channels, so it is clear
that the CSA and NNCC methods allow one to handle much larger problems
than the full CC method. Even more significant is the CPU time for the
propagation, which depends roughly on the third power of the number of
channels. Our renormalized Numerov propagation involved 501 steps and
took 6633 CPU seconds for the CC method, 0.40 seconds for CSA, and 4.15
seconds for NNCC on a single Intel Xeon Platinum 8268 processor. So even
with the more advanced NNCC approximation the savings in CPU time
relative to full CC calculations is substantial. We note here that the
actual gain in CPU time and matrix size in CSA and NNCC with respect to
full CC calculations depends on the characteristics of the channel basis.
In our calculations of the cross sections in rovibrationally
inelastic CO$_2$-He collisions, we had to use a channel
basis with large maximum values of the monomer rotational angular
momentum $j$ and of the total angular momentum $J$.

In Ref.~\cite{selim:21} we explained that the MC-DWBA method applied in
full CC calculations reduces the CPU time by about a factor of 3. Here
we find that also when combined with the NNCC method MC-DWBA leads to a
further decrease of CPU time by a factor of 3.

\section{Conclusions} \label{sec:concl} The advanced models currently
being developed by astronomers \cite{bosman:17,bosman:19} and the
availability of data from the James Webb space telescope (JWST) in the
near future require the knowledge of rovibrational state-to-state
collisional rate coefficients. These rate coefficients can be obtained
from coupled-channel (CC) scattering calculations, but these are very
demanding. Here we presented more efficient methods based on the
coupled-states approximation (CSA) in which one neglects the
off-diagonal Coriolis coupling in the scattering Hamiltonian in
body-fixed coordinates. This makes $\Omega$, the projection of the total
angular momentum $J$ on the intermolecular axis, a good quantum number,
so that scattering calculations can be performed independently for each
$\Omega$. In addition to CSA, we investigated a method called NNCC
(nearest-neighbor Coriolis coupling) \cite{yang:18} that includes
Coriolis coupling to first order by simultaneously including basis
functions with $\Omega$ and $\Omega \pm 1$. The NNCC method is more
expensive than the CSA method, but still much more efficient than full
CC calculations. We tested three versions of the CSA method and two
versions of the NNCC method. The cross sections and rate coefficients
from the two NNCC methods are similar, and substantially better than all
CSA results.

All of this is illustrated by showing state-to-state cross sections and
rate coefficients of rovibrational transitions induced in CO$_2$ by
collisions with He atoms. Results from the CSA and NNCC methods are
compared in detail with those from full CC calculations. In a recent
paper \cite{selim:21} we have shown that the application of the
multi-channel distorted-wave Born approximation (MC-DWBA) in CC
calculations reduces the required CPU time by a factor of 3. Here we
show that a further increase of effiency by about the same factor can be
obtained by applying MC-DWBA to the NNCC method, with practically no
loss of accuracy.

Finally we note that rovibrational transitions in CO$_2$ are probably
more strongly induced by collisions with H$_2$ than by collisions with
He, for several reasons. First, CO$_2$-H$_2$ interactions are stronger
than CO$_2$-He interactions because H$_2$ is more polarizable than He
and it has a quadrupole moment. Secondly, as was recently shown for
rovibrationally inelastic H$_2$O-H$_2$ collisions \cite{wiesenfeld:21},
such transitions may be enhanced by simultaneous rotational excitation
of H$_2$. The channel basis needed in CO$_2$-H$_2$ scattering
calculations will be even larger than the bases we used for CO$_2$-He
and the efficient methods to compute rovibrationally inelastic collision
cross sections and rate coefficients presented here will be very
advantageous.

\section*{Data availability}
The data that supports the findings of this study are available within
the article.

\section*{Acknowledgements}
We thank Ewine van Dishoeck and Arthur Bosman for stimulating and useful
discussions. The work is supported by The Netherlands Organisation for
Scientific Research, NWO, through the Dutch Astrochemistry Network
DAN-II. We also acknowledge a useful discussion with David Manolopoulos.

\clearpage


\end{document}